\documentstyle[aps,prl,twocolumn,epsf]{revtex}
\begin{document}

\twocolumn[\hsize\textwidth\columnwidth\hsize\csname @twocolumnfalse\endcsname

\draft

\title{ Spectral functions in a magnetic field as a probe of spin-charge 
separation in a Luttinger liquid}

\author{Silvio Rabello and Qimiao Si}

\address{Department of Physics, Rice University, Houston, TX 77005-1892, USA}

\maketitle
\begin{abstract}

We show that the single-particle spectral functions in a magnetic
field can be used to probe spin-charge separation of a Luttinger liquid. 
Away from the Fermi momentum, the magnetic field splits both the spinon
peak and holon peak; here the spin-charge separation nature is reflected
in the different magnitude of the two splittings. At the Fermi momentum,
the magnetic field splits the zero-field peak into {\it four} peaks.
The feasibility of experimentally studying this effect 
is discussed.

\end{abstract}
]

Spin-charge separation is a clear-cut example of what could happen in
a non-Fermi liquid metal. The defining characteristics of spin-charge
separation are two-fold. First, there are  more than one type of 
elementary excitations. Second, and equally important, one kind of
excitations carries spin quantum number only while the other
carries charge quantum number only.
Theoretically, this phenomenon is well-established in one
dimension\cite{Lieb-Wu,Voit-rev}. Whether, and how, it occurs in two 
dimensions remains an active topic of current studies. 

Our goal in this work
is to seek for
experimental manifestations of spin-charge separation\cite{spin-transport}. 
We demonstrate that the single-particle spectral functions in a magnetic
field can be used to probe spin-charge separation.
For concreteness, we will focus on the one-dimensional Luttinger
liquid. In general, the single-particle spectral function of 
a Luttinger liquid is expected to contain two dispersive peaks,
at the spinon and holon energies respectively\cite{Voit,Meden,Ren}.
Phenomenologically, to establish that the dispersive features of 
the spectral function indeed correspond to spinon and holon peaks
instead of, say, simply two ordinary electron bands, it is necessary
to determine the quantum numbers of these excitations.
In this paper, we show that a Zeeman coupling can be used for this 
purpose.

The natural language to describe  the Luttinger model is the bosonization
of the electronic degrees of freedom\cite{Delft}.
We write the boson representation of the fermion fields as follows,
\begin{eqnarray}
\Psi_{r\sigma} (x)=\lim_{a\rightarrow 0}\frac{ e^{irk_F x}}{\sqrt {2\pi a}}
F_{r\sigma} e^{ir\Phi_{r\sigma}(x)} 
\label{fermion}
\end{eqnarray}
$\Psi_{r\sigma}$ describes fermions with 
spin $\sigma=\uparrow,\downarrow$ on two branches ($r=\pm 1$) with linear dispersion [$\varepsilon_r(k)= v_F (rk - k_F)$] about the two Fermi points $\pm k_F$.The boson field $\Phi(x)$ is defined as:
\begin{eqnarray}
\Phi_{r\sigma}(x) =
&& 
{2 \pi x \over L} N_{r\sigma}
+\sum_{q>0} \sqrt{2\pi \over q L } (-i b_{qr\sigma}^\dagger 
e^{iqx} 
\nonumber\\&&
+ i b_{qr\sigma} e^{-iqx})e^{-qa/2}
\label{bos4}
\end{eqnarray}
where the Tomonaga bosons are related to the original electron operators by
$b_{qr\sigma} = \sqrt{2 \pi/q L} 
\sum_k \Psi_{kr\sigma}^\dagger\Psi_{k+q~r\sigma}$,
and 
$x \in [-L/2, L/2]$. 
$N_{r\sigma}$ and $F_{r\sigma}$ represent the zero
modes:
$N_{r\sigma}$ is the deviation of the
conduction electron occupation number from the chosen reference state value,
while the Klein factor $F_{r\sigma}^{\dagger}$
($F_{r\sigma}$) raises (lowers) $N_{r\sigma}$ by one.

In terms of the boson variables 
$\Phi_{c,s}=1/\sqrt{2}(\Phi_{\uparrow}\pm\Phi_{\downarrow})$,
the zero field Luttinger Hamiltonian assumes the simple
charge-spin separated form:
\begin{eqnarray}
\label{Ham}
H=\sum_{\nu=c,s}\frac{v_\nu}{2\pi}\int ~dx \left( K_\nu(\partial_x\theta_\nu)^2+\frac{1}{ K_\nu }(\partial_x\phi_\nu)^2\right) \;\; ,
\end{eqnarray}
with $\theta_\nu,\phi_\nu$ defined through $\Phi_{r,\nu}=\phi_\nu+r\theta_\nu$. The charge and spin velocities are given in terms of the original 
Fermi velocity and the 
interaction strengths $g_{ic,s}$
\begin{equation}
\label{velo}
v_{c,s}=\sqrt{\left(v_F+\frac{g_{4c,s}}{\pi}\right)^2-\left(\frac{g_{2c,s}}{\pi}\right)^2}\;\;.
\end{equation}
The stiffness constants $K_{c,s}$ are given by 
\begin{equation}
\label{K}
K_{c,s}=\sqrt{\frac{\pi v_F+g_{4c,s}-g_{2c,s}}{\pi v_F+g_{4c,s}+g_{2c,s}}}\;\;.
\end{equation}
Here $g_{ic,s}=\frac{1}{2}(g_{i\|}\pm g_{i\perp})$, for i=2,4
are interactions between density fluctuations at the same or opposite
Fermi points respectively 
and $\|,\perp$ refer to parallel or anti-parallel spins.
We have here assumed a simplified model where 
the couplings $g_i$ are momentum independent.
For a spin-rotationally invariant system  at the fixed point,
$K^*_s$ is equal to unity and the backscattering term is
renormalized to zero. Additional terms induced by the curvature of 
the band dispersion are not included in Eq. (\ref{Ham}), since they are 
irrelevant in the renormalization group (RG) sense\cite{Voit-rev}.

We now turn a magnetic field on  by  adding a Zeeman term to 
the Hamiltonian,
\begin{equation}
\label{hterm}
H_h=-h\sum_{r,\sigma}\sigma N_{r\sigma}\;.
\end{equation}
For small fields, the renormalized parameter
\begin{equation}
\label{K*}
K_s^*\simeq 1+\frac{1}{2ln(h_c/h)}\;\; ,
\end{equation}
with $h_c\simeq v_sk_F$ being the critical field that spin polarizes
the sample. 
Eq. (\ref{K*}) agrees with  the  Bethe-{\it Ansatz} results for
the 1D positive U Hubbard model in a magnetic field \cite{FK} \cite{PS}.
The magnetic field affects the single particle spectrum through
the zero modes. The first effect is through the time dependence
of the Klein factors, 
\begin{eqnarray}
\label{Ft} 
F_{r\sigma}(t)
=
&=&
F_{r\sigma}(0)
exp(-i\sigma\{\frac{\pi v_s}{2L}[K_s(N_{r,s}-N_{-r,s})
\nonumber\\&&
+\frac{1}{K_s}(N_{r,s}+N_{-r,s})]-h\}t)\; .
\end{eqnarray}
However, to the linear order in $h$, the expectation value 
of $N_{r,s}=N_{r\uparrow}-N_{r\downarrow}$ is just equal 
to $L\chi h/2$, with $\chi =2K_s(\pi v_s)^{-1}$. 
So when computing Green's functions of $\Psi$ we can set 
$F_{r\sigma}$ to be time independent,
\begin{equation}
\label{Klein}
F_{r\sigma}(t)=F_{r\sigma}(0)\;\; . 
\end{equation}
The remaining effect is that $\Phi_{r\sigma}(x)$ 
acquires the ground state expectation value, 
$\langle\Phi_{r\sigma}(x)\rangle=\sigma x\frac{K_s}{v_s} h
\simeq\sigma x\frac{h}{v_s}$.
This in turn corresponds to a splitting in  $k_F$  
\begin{equation}
\label{shift}
k_{F\sigma}=k_F+\sigma\frac{h}{v_s}\;\;. 
\end{equation}

We  can now discuss the fate of spin-charge separation in a magnetic
field. 
For models with a quadratic dispersion, $v_{F\uparrow}\neq v_{F\downarrow}$.
To the linear order in $h$, 
\begin{equation}
\label{mix}
\Delta v_F=v_{F\uparrow}-v_{F\downarrow}=2v_F\frac{h}{h_c}
\end{equation}
This leads to a mixing of spin and charge variables,
\begin{equation}
\label{mixH}
H_{mix}=\frac{\Delta v_F}{2\pi}\int ~dx \left( \partial_x\theta_c\partial_x\theta_s
+ \partial_x\phi_c\partial_x\phi_s\right)
\end{equation}
For all relevant fields and band fillings $\Delta v_F$ is very small.
We can diagonalize $H$ by going to new variables
$(\phi_{c,s}',\theta_{c,s}')$. The new velocities $v_c'$ and $v_s'$
differ from $v_c$ and $v_s$ respectively only to order 
$(\Delta v_F)^2$. Namely, to the linear order in $h$,
$v_c'=v_c$ and $v_s'=v_s$;
the mixing will be visible only as a correction to the critical
exponents in correlation functions (see below).
In this sense, spin-charge separation is preserved
to the linear order in $h$.

The survival of spin-charge separation allows us to understand 
the physical meaning of Eqs. (\ref{Klein},\ref{shift}).
By introducing Klein factors for spinon and holon\cite{Silvio2},
we can see that to the linear order in $h$ the spinon ``Fermi momentum''
is shifted to $k_F+\sigma h/v_s$ while the holon ``Fermi momentum'' remains
at $2k_F$. The Fermi energy is also unaffected to this order.

We now calculate the 
total single-electron spectral function:
\begin{eqnarray}
\label{Lehm1} 
A(q,\omega)
=
\sum_{\sigma}
A_{\sigma}(q,\omega)
\end{eqnarray}
We will measure momentum with respect to the zero-field Fermi momentum
($q \equiv k - k_F$) and energy with respect to the (field-independent)
Fermi energy $E_F$. $A_{\sigma}(q,\omega)$ is determined by
the  imaginary part of the Fourier transform of the
retarded electron Green's function:
\begin{eqnarray}
\label{green} 
G^R_{r,\sigma}(x,t)&=&-i\theta(t)
\langle 0|\{\Psi_{r,\sigma}(x,t),\Psi^\dagger_{r,\sigma}(0,0)\}|0\rangle
\nonumber\\
&=& \theta(t)\frac{e^{irk_{F\sigma}}}{2 \pi i}\{\prod_{\nu = c,s}
\left(\frac{a^2}{(a+iv_\nu t)^2+x^2}\right)^{2\beta_\nu^\sigma\gamma_\nu}
\nonumber\\
&&\times 
\frac{1}{[a +i(v_{\nu}t-rx)]^{\beta_\nu^\sigma}}+
(x,t\rightarrow -x,-t)\}
\nonumber\; ,\\ 
\end{eqnarray}
where to linear order in $\Delta v_F$ and $\Delta v_F<< v_c-v_s$ the  exponents are 
\begin{equation}
\label{beta}
\beta_{c,s}^\sigma=\frac{1}{2}(1\pm \sigma\frac{v_c(K_c+K_c^{-1})+v_s(K_s+K_s^{-1})}{2(v_c^2-v_s^2)}\Delta v_F)
\end{equation}
and
\begin{equation}
\label{gamma}
\gamma_{\nu}=\frac{1}{8}(K_\nu+\frac{1}{K_\nu}-2)\; .
\end{equation}

\begin{figure*}[h]
\centerline{\epsfysize=5cm\epsfbox{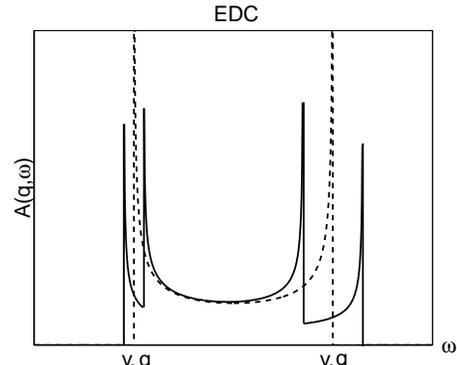}}
\nonumber
\caption{Energy distribution curve (EDC) of the spectral function $A(q,\omega)$
for a given $q\equiv k-k_F>0$. The solid (dashed) curve corresponds to
the finite (zero) field case.
}
\end{figure*}

We first consider the Luttinger model with only  $g_{4\perp}\neq 0$, which 
corresponds to the one-branch Luttinger model where there is no
communication between right and left movers  but still  is  spin-charge 
separated with corresponding velocities  $v_{c,s}=v_F\pm g_{4\perp}/{2\pi}$.
Although this is a simplified model it captures the 
basic physics of the electron decay into charge and spin collective
excitations. (For zero field case see \cite{Voit}.)
We restrict our attention to 
the $r=+$ branch, and to the electron injection process:
\begin{eqnarray}
\label{spec1}
&&A(q,\omega)\sim\sum_\sigma\int_{\sigma h/v_s}^{q}dk_s \frac{\delta(\omega-\varepsilon_c(q-k_s)-\varepsilon_s(k_s)+\sigma h)}
{|\varepsilon_c(q-k_s)|^{\beta_c^\sigma}|\varepsilon_s(k_s)-\sigma h|^{\beta_s^\sigma}} , \nonumber
\\
&&
\end{eqnarray}
At $h=0$ the spectral function 
has a continuum with well defined edges and power law
singularities (Fig.1). The edges trace out the spin and charge dispersion
relation, $\varepsilon_{c,s}(q)\equiv v_{c,s}k$.
The lower (spinon) edge corresponds 
to where all the electron momentum is carried by the spinon
and the upper (holon) edge where the anti-holon carries all q. 
 
When $h\neq 0$ the region of nonzero spectral weight is in between the 
frequencies $v_s(k-k_{F\uparrow})<\omega<v_c(k-k_{F\downarrow})$ 
$(v_c>v_s)$. 
We assume a small $\Delta v_F$ and keep
the exponents of both singularities close to $1/2$.
The spin and charge edges  are respectively split by 
\begin{eqnarray}
\label{spinon}
\Delta\omega_s&=&2 h \\
\label{holon}
\Delta\omega_c&=&2\frac{v_c}{v_s} h\;\; .
\end{eqnarray}
The resulting energy distribution curve (EDC) is shown in Fig. 1.
That the holon peak is also split by a magnetic field,
while surprising at the first sight, can be understood as follows:
As a holon is knocked out
it is always accompanied by a spinon whose energy
is shifted. What is not obvious is that the holon peak is split
by a magnitude different from that of the splitting of the
spinon peak. This is due to the fact that the splitting by the
magnetic field takes place in $k-$space:
The magnetic field splits the ``spinon Fermi surface'' by 
$\Delta k_{s,\sigma}=\sigma h/v_s$, without
changing the ``holon Fermi surface''. 
Since the single-electron Green's function is a convolution of the 
spinon and holon Green's functions, the energy 
change for the  spinon peak is then $v_s\Delta k_{s,\sigma}$, while
that for the holon peak is $v_c\Delta k_{s,\sigma}$. 

The field effect on the momentum distribution curve (MDC) is very
different. When the reference point of the spinon momentum
changes, both edges respond equally in the MDC. As a result,
both peaks are split by the same amount $2h/v_s$, as 
illustrated in Fig. 2.

\begin{figure*}[h]
\centerline{\epsfysize=5cm\epsfbox{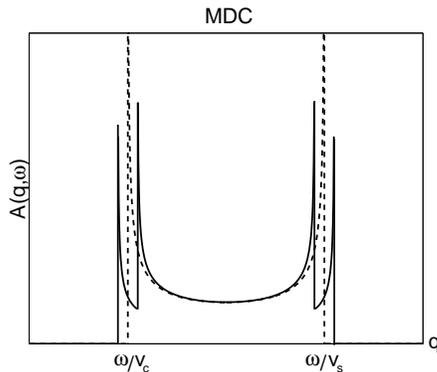}}
\nonumber
\caption{Momentum distribution curve (MDC) of the spectral function
for a given $\omega\equiv E-E_F>0$.
The solid (dashed) curve corresponds to the finite (zero)
field result.
}
\end{figure*}

The difference in the splitting of the spinon  and 
holon edges in the EDC  provides a means to determine the spin quantum 
number of the two excitations.
Suppose we divide the Zeeman energy between the holon and  spinon in 
a parametrized fashion, i.e. $(1-\lambda)h$ for the spinon and 
$\lambda h$ for the anti-holon with $\lambda\in [0,1]$, 
from (\ref{spec1}) we have 
\begin{eqnarray}
\label{rel}
\frac{\Delta\omega_c}{\Delta\omega_s}&=&\frac{v_c}{v_s} \\
\Delta\omega_c -\Delta\omega_s&=&2 (\frac{v_c}{v_s}-1)(1-\lambda+
\frac{v_s}{v_c}\lambda)h\;\; .
\end{eqnarray}
That tells us that if the edge splittings (\ref{spinon}) and (\ref{holon}) 
are observed then $\lambda=0$ and all the magnetic coupling  is carried
by the spinon.

Let's now turn to more realistic models.
A finite  $\gamma_c$
couples left to right movers and this introduces anomalous 
fermion exponents for the edge singularities and  generates 
spectral weight in Fig. 1 for frequencies above the $v_c q$ edge  
and a small cusp below $-\varepsilon_c(q)$ \cite{Voit}. The 
calculation of the spectral function in this case is rather involved 
but the power law behavior at the spinon and holon edges can be 
extracted easily by power counting:
\begin{eqnarray}
\label{powlaw} 
A_\sigma(q,\omega\sim \varepsilon_s(q))&\sim& 
|\omega-\varepsilon_s(q)+\sigma h|^{-\alpha_s^{\sigma}}
\\
A_\sigma(q,\omega\sim \varepsilon_c(q))&\sim& 
|\omega-\varepsilon_c(q)+\frac{v_c}{v_s}\sigma h|
^{-\alpha_c^{\sigma}}
\end{eqnarray}
Where $\alpha_s^{\sigma} = \beta_c^\sigma
-2\gamma^\sigma_c-\gamma^\sigma_s $,
$\alpha_c^{\sigma}
= \beta_s^\sigma -
2\gamma^\sigma_s - \gamma^\sigma_c$,
with $\gamma^\sigma_\nu=2\beta_c^\sigma\gamma_\nu$.
The peak structure for both EDC at $q\ne 0$ and MDC at $\omega \ne 0$
remains essentially unchanged from Figs. 1 and 2.

Consider now the EDC at the Fermi momentum $k=k_F$. 
For the zero field case it has a power law singularity determined by 
the anomalous exponent $\gamma_c$ and no sign of spin-charge separation
can be seen. Turning on a finite $h$ now splits this peak into {\it four}
peaks, as seen in Fig. 3.
Here even at $k_F$ and fairly strong coupling ($\gamma_c=0.2$, 
$\gamma_s\sim 0$) the sign of spin-charge separation is still visible
as the Zeeman lines are split into contributions coming from the spinon
and holon edges. As $\gamma_c$ increases we enter the strong coupling
regime and the edge singularities
are more and more obscured as the anomalous exponent dominates over 
the spin-charge separated character of the single particle spectrum.  

\begin{figure}[h]
\centerline{\epsfysize=5cm
\epsfbox{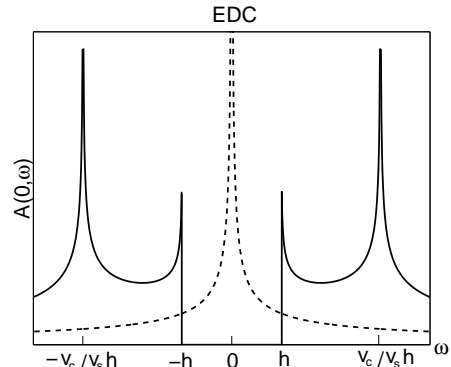}
}
\caption{EDC at $k=k_F$ for $\gamma_c=0.2$ and
$\gamma_s\sim 0$ (dash curve for $h=0$).}
\end{figure}

This splitting of one peak at Fermi momentum into four represents the most
dramatic and direct manifestation of electron fractionalization. 
It unambiguously shows that the initial electron peak is in fact a composite
of two different elementary excitations.

We turn next to MDC at the Fermi energy $E_F$.
In the $h=0$ case it is,
for $\gamma_c=0$, just the delta function $\delta(q)$.
For $h\neq 0$ it is easy to verify, from the Lehmann representation
of $A(q,\omega)$
and  the properties of $\Psi_{r,\sigma}(x)$ zero modes in a magnetic
field, that 
\begin{equation}
\label{mdc0}
A(q,0)\sim \sum_{\sigma}\delta (q-\sigma \frac{h}{v_s})\;.
\end{equation}  
The only interaction effect that remains in this MDC is the renormalization of 
$v_F$ to $v_s$ in the Zeeman splitting of the Fermi momentum. 
(A finite $\gamma_c$ would turn 
each delta function
into a peak whose width
and height depend on temperature in a power-law fashion.)

We stress that the contrasting behavior of the Zeeman-splittings
in EDC and MDC reflects a generic feature of spin-charge separation.
Namely, the main effect of the  magnetic field is to split the spinon
Fermi momentum.

Finally, the integrated spectral weight $N(\omega)\sim 
|\omega|^\alpha$, with $\alpha =2(\gamma^\sigma_c+\gamma^\sigma_s)$.
Here aside from the exponents the magnetic field has no other effect.

In several quasi-one-dimensoinal materials,
ARPES experiments have seen two dispersive peaks\cite{expts1D,Voit-rev2}.
One interpretation is that these two peaks correspond to dispersing
spinon and holon modes respectively.
In the high $T_c$ cuprates, with fastly improving resolution in both
energy and momentum, ARPES is now providing both EDC and
MDC\cite{Johnson}.
The theoretical interpretation
of the lineshapes is actively being
pursued\cite{MFL}.
Our results imply that studying the Zeeman effect on
the spectral functions
can provide valuable information about the quantum numbers
of the elementary excitations. 

We conclude with a few remarks concerning the experimental
implementations. 
The Zeeman effects can in principle be studied using ARPES
in a magnetic field. 
For 1D systems, an alternative technique to study the magnetic field 
effect is the momentum-resolved tunneling\cite{Grigera,AJS2}.
We illustrate the quantitative effects of the magnetic field on the
exponents by using seminconductor quantum wires as an example.
We take\cite{Auslaender} $K_c \approx 0.7$ and 
$E_{F} \approx 20 meV$. For a vanishing field, 
$\alpha_s^{\sigma} \approx 0.47$ and 
$\alpha_c^{\sigma} \approx 0.48$.
For a field of, say, $10$ T, and using $g \approx 0.4$ for GaAs,
we estimate the corrections to
$\alpha_s^{\sigma}$ and $\alpha_c^{\sigma}$ 
to be small, of the order of $\pm 0.05$.
The experiments have to be done at temperatures below 
the Zeeman splitting, which are relatively easy to access.
We also note that the situation would be even better for materials
with high $g$ factors, such as InSb for which $g\sim -50$ \cite{Made}.

We would like to thank G. Aeppli, N. Andrei, P.-A. Bares, S. A. Grigera,
A. J. Schofield, and especially R. Gatt, for useful discussions. 
This work has been supported by the Robert A. Welch Foundation,
NSF Grant NSF Grant No.\ DMR-0090071, and TCSUH.

\end{document}